\documentstyle[aps,prb,epsfig]{revtex}
\begin{document}
\draft

\title{Superconductivity in the Extended Hubbard Model with More Than  
Nearest-Neighbour Contributions}

\author{Zsolt~Szab\'o and Zsolt~Gul\'acsi}
\address{Department of Theoretical Physics, Lajos Kossuth University,\\
H-4010 Debrecen, P.O. Box 5., Hungary}

\date{\today}

\maketitle

\begin{abstract}
Superconducting phase diagram of the extended Hubbard model supplemented
with interaction and hopping terms exceeding nearest neighbour distance in 
range is analysed systematically at different band-filling and 
temperature values in a mean-field approximation. The obtained results clearly
underline the importance of next-nearest neighbour terms in developing 
the main superconducting properties of the model system. In particular, the 
emergence of superconducting phases of different symmetry at a 
given point of the phase diagram, critical temperatures $T_c$, zero 
temperature gap amplitude values $\Delta_{0}$, $\Delta_{0}/T_{c}$ ratios, 
doping and temperature dependences are all strongly influenced by 
next-nearest neighbour contributions.
\end{abstract}

\section{Introduction}

The subject of high  critical temperature superconductors starting from an 
early proposal of Anderson (Anderson, 1987) polarizes significant 
forces around the Hubbard model, its extensions and derivatives 
(Dagotto, 1994). On this line, taking the fact into account that high 
temperature superconductivity generally emerges in 
systems with a pronounced planar structure (Smith, Manthiram, Zhou, 
Goodeneough and Market, 1991), the two-dimensional square lattice
version of the Hubbard model in its extended version is 
intensively used as a starting point for describing superconducting
properties of this type of materials (Micnas, Ranninger and
Robaszkiewicz, 1990). As a consequence, within the last few years the Hubbard
Hamiltonian containing nearest neighbour ({\em NN}) interaction and hopping 
terms, has been extensively studied (de Boer, de Chatel, Franse and de
Visser, 1995) and  the 
emergence of superconductivity has been rigorously demonstrated in some 
conditions (de Boer, Korepin and Schadschneider, 1995) in these
systems. In the last period, however, a great number of results
directed our attention to the importance of 
extending the interaction and hopping range in such type of an analysis. The 
stimulating observations can be summarized as follows.

Concerning the kinetic energy term, Koma and Tasaki starting from the 
approximation-free study of two-point correlation functions at finite 
temperatures in low dimensional extended Hubbard models containing
more than {\em NN} hopping terms (Koma and Tasaki, 1992) sharply 
demonstrated the fact 
that electron hopping with higher than {\em NN} terms plays a 
fundamental role in various condensation phenomena in itinerant-electron
systems. The importance of hopping-range in building up the
phase diagram was also emphasized in an exact manner for finite- 
(Tasaki, 1995) and infinite-{\em U} (Verges, Guinea, Galan, van
Dongen, Chiappe and Louis, 1994) Hubbard models. Furthermore, as it was 
pointed out by Veilleux {\em et al.} (Veilleux, Dar\'e, Chen, Vilk and
Tremblay, 1995) within a model containing 
only {\em NN} hopping the Fermi surface topology and the filling dependence 
of both the Hall coefficient and the uniform magnetic susceptibility are 
qualitatively wrong, a disagreement that cannot be removed perturbatively. On
the other hand, incorporating also next-nearest neighbour ({\em NNN}) 
hopping the band structure becomes more realistic and all the above mentioned
physical quantities, as well as the position of neutron scattering intensity 
maxima, have the correct qualitative behaviour (Lavagna and Stemmann,
1994, Littlewood, Zaanen, Aeppli and Monien, 1994). Following this line of 
reasoning it is important to mention that the relative influence of density of
state (DOS) effects - like van Hove singularities (Blumberg,
Stojkovic and Klein, 1995) - and Fermi 
surface topology effects - like nesting (Gul\'acsi, Bishop and
Gul\'acsi, 1995) - on pairing are extremely 
important in describing superconducting properties of high-$T_c$ materials. 
While both van Hove singularity and nesting occur simultaneously at half 
filling in the usual {\em NN} model, there is no nesting when {\em NNN} 
hopping is present (Veilleux {\em et al.} and references cited therein) and 
the influence of saddle point effects (Abrikosov, 1995, Abrikosov,
1994) can be taken more accurately into consideration. Furthermore,
the presence of hopping exceeding 
{\em NN} terms in range was observed, in particular, by thermopower 
measurements in related systems (Ponnambalam and Varadaraju, 1995) and
was found to be consistent with angle-resolved photoemission data 
(Fehrenbacher and Norman, 1995) and gap symmetry 
analysis (O'Donovan and Carbotte, 1995) too. We also would like to
mention that the ratio of {\em NN} to {\em NNN} hopping amplitudes is 
clearly known for a broad spectrum of high-$T_c$ materials (Brenig, 1995).

Concerning the interaction terms, it is known from exact studies of
phase diagrams related to extended Hubbard models in higher than one
dimension (Strack and  Vollhardt, 1994) that contributions exceeding
{\em NN} distances 
in range can significantly influence the emergence of condensed phases. 
Up to this moment such type of studies have been carried out on magnetic 
phases, however, the relevance of the obtained results to superconductivity 
is quite straightforward (Strack and Vollhardt, 1995). Similarly, the
non-negligible character of {\em NNN} interaction contributions was 
clearly pointed out experimentally. For example, the importance of
{\em NNN} type off-site 
interactions was revealed in the interpretation of Auger core-valence line 
shapes (Verdozzi and Cini, 1995), which allows the direct observation
and measurement of 
interparticle interaction strength and its distance dependence in valence 
bands of solids (Verdozzi, 1993). Furthermore, King {\em et al.}
based on angle resolved photoemission data (King {\em et al.}, 1993), 
clearly underline that model 
Hamiltonians containing interaction terms only up to the {\em NN} ones 
are not sufficient for the proper description of the electronic structure 
near the Fermi energy ($E_F$) in {\em NCCO} type materials. They also 
conlude that {\em NNN} interactions are {\em very} important and models 
including only {\em NN} interactions in $CuO_2$ planes are unable to 
explain related clear experimental facts. The interpretation of 
two-magnon Raman scattering in high-$T_c$ materials (specially the $A_{1g}$ 
and $B_{2g}$ finite intensity peaks) also requires second-nearest neighbour or 
even longer-range contributions to be incorporated at the level of the
interaction terms (Brenig). Moreover, the importance of {\em NNN} terms
in developing basic properties of extended Hubbard models has been 
claimed (Di Stasio and Zotos, 1995), their effect on screening
processes and as well as on phase diagram configurations has been
studied (van den Brink, Meinders, Lorenzana, Eder and Sawatzky, 1995),
the relevance to {\em YBCO} materials has been emphasized for
various conditions (Grigelionis, Tornau and
Rosengren, 1996) and their importance in loss 
of antiferromagnetic order has been discussed (Kampf, 1994). 

Stimulated by these findings we present an extensive study of superconducting
phase diagram of the extended Hubbard model containing more than 
nearest neighbour contributions, in order to provide a systematic image about 
the importance of the longer range contributions. The study is given in 
two dimensions, based on a traditional Hartree-Fock type 
decoupling procedure in terms of the Green's function description allowing, 
however, arbitrary gap symmetry. The DOS is treated exactly, 
and the stable superconducting state is chosen on the basis of a
free-energy type analysis. Such type of a {\em systematic} investigation, 
incorporating the effects of {\em NNN} terms too, has not been
presented so far in literature.

Our results emphasize the importance of interaction and
hopping terms {\em exceeding} nearest neighbouring distance in range
in building up 
superconducting characteristics of the system described by an extended Hubbard
type model. In particular, the emergence of superconducting phases 
of different symmetry at a given point of the phase diagram, critical
temperatures, zero temperature gap amplitude values, $\Delta_0/T_c$ ratios,
doping and temperature dependences are strongly influenced by the presence of
{\em NNN} terms. Based on the results we conclude that the effect of 
{\em NNN} terms is not marginal but essential in the description of basic 
superconducting properties.
 
The paper is organized as follows. Section \ref{themodel} presents the model, 
Sec.\ \ref{results} contains besides the numerical procedure 
(Sec.\ \ref{numproc}) the obtained results grouped into 
three paragraphs as follows: Sec.\ \ref{NNTB} analyzes the {\em NN} case, 
Sec.\ \ref{NNTBF} outlines the effect of hopping terms exceeding {\em NN} 
distances in range and Sec.\ \ref{NNNTBF} presents the effects of {\em NNN} 
interaction contributions, the study being given at various temperature and
doping values. Finally, the conclusions and summary close 
the presentation in Sec.\ \ref{concl}.

\section{The model and its description}
\label{themodel}

In this paper the following extended Hubbard Hamiltonian is considered on a
square lattice:
\begin{eqnarray}
\label{hamiltonian}
H &=& - \sum_{i,j} \: \sum_{\sigma} \: t_{ij} \: c_{i\sigma}^{\dagger} \: 
c_{j\sigma} \: + \: \frac{1}{2} \: \sum_{i,j} \: \sum_{\sigma,\sigma'} \:
U^{\sigma\sigma'}_{ij} \: n_{i\sigma} \: n_{j\sigma'} \;,
\end{eqnarray}
where the fermion operators $c_{i\sigma}^{\dagger}$ and
$c_{i\sigma}$ create and annihilate, respectively, electrons with spin
$\sigma$ in the single tight-binding orbital associated with site {\em i}, and
$n_{i\sigma}$ is the particle number operator. The parameters 
$t_{l} \!\equiv\! t_{ij}$ with $l \!=\! 1,2,\ldots,5$ are the real space 
hopping matrix elements between sites {\em i} and {\em j} with a diagonal part 
$t_{0} \!\equiv\! t_{ii}$. The coupling constants $U^{\sigma\sigma'}_{ij}$ 
are independent parameters denoted below as 
$U_{0} \!\equiv\! U^{\sigma,-\sigma}_{ii}$, 
$U_{1} \!\equiv\! U^{\sigma\sigma'}_{i,i+1}$ and 
$U_{2} \!\equiv\! U^{\sigma\sigma'}_{i,i+2}$ 
on-site, nearest neighbour and next-nearest neighbour contributions, 
respectively. All higher order interaction terms (for $j \!>\!i\!+\!2$) 
will be neglected for the present study, that is, 
$U_{l} \!\equiv\! U^{\sigma\sigma'}_{i,j>i+2} \!=\! 0$. Furthermore, any of 
$U_{0}$, $U_{1}$ and $U_{2}$ can either be positive or negative with no further 
restriction. This general choice of interaction constants allows us a 
systematic exploration of the full superconducting phase diagram of the 
model under study. 

The gap equation obtained within a Hartree-Fock decoupling 
procedure (Abrikosov, Gor'kov and Dzyaloshinskii, 1965) is given by 
\begin{eqnarray}
\label{opequation}
\Delta(\vec{k}) &=& - \: \sum_{\vec{q}} \: U(\vec{q}-\vec{k}) \:
\Delta(\vec{q})
\: F(\vec{q}) \;,   
\end{eqnarray}
and the chemical potential $\mu$ is controlled via 
\begin{eqnarray}
\label{muequation}
1 \: - \: \delta &=& \sum_{\vec{k}} \: \left( \: 1 \: + \: \Omega(\vec{k}) \:
F(\vec{k}) \: \right) \;.
\end{eqnarray}
Here $\delta \!=\! 1\!-\!2n$ is the band filling factor with positive or 
negative values corresponding to hole or electron doping, respectively.
Furthermore, $n$ is the particle number density for a given spin index. 
In the above equations  
\begin{eqnarray}
F(\vec{k}) &=& \frac{\tanh(\: \frac{1}{2} \: \beta \: E(\vec{k}))}{2 \: 
E(\vec{k})} \;, 
\\
E(\vec{k}) &=& \sqrt{\: \Omega^{2}(\vec{k}) \: + \: \Delta^{2}(\vec{k})} \;, 
\\
\Omega(\vec{k}) &=& \mu \: - \: \epsilon(\vec{k}) \: - \: n \: \overline{U}, 
\label{equss}
\end{eqnarray}
with $\overline{U} \: = \: U_{0} \:+ \: 8U_{1} \: + \: 8U_{2}$, and $\beta$ 
stands for the reciprocal temperature. Taking the lattice spacing to be unity,
the explicit expression of the inter-particle coupling in 
{\bf k}-space for a square lattice is given by
\begin{eqnarray}
\label{pairpotential}
U(\vec{k}) &=& U_{0} \: + \: 2U_{1} \: (\cos k_{x} \: + 
\: \cos k_{y}  ) \: + \: 4U_{2} \: \cos k_{x} \cos k_{y}  \;.
\end{eqnarray} 
One can easily verify that in Eq.\ (\ref{opequation}) 
\begin{eqnarray}
\label{separpot}
U(\vec{q}-\vec{k}) &=& \sum_{i \: \in \: {\cal R}} \: g_{i} \: 
\Psi_{i}(\vec{k}) \: \Psi_{i}(\vec{q}) \;,
\end{eqnarray}
where $\cal R$ collects all the irreducible representations of the point
group ($C_{4v}$) characterizing the square lattice, $\Psi_{i}(\vec{k})$ 
represents the basis functions of the i{\em th} irreducible representation 
and $g_{i}$ denotes the effective coupling constants 
(see Table\ \ref{tablazat}). As a consequence, the {\bf k}-dependent
order parameter can be given  in terms of these basis functions, that is,
\begin{eqnarray}
\Delta(\vec{k}) &=& \sum_{i \: \in \: {\cal R}} \: \Delta_{i} \: 
\Psi_{i}(\vec{k}) \;.
\label{ez2}
\end{eqnarray} 
In Eq.\ (\ref{ez2}) {\em i} represents both singlet and triplet pairing. 
Based on Eqs.\ (\ref{opequation}) and (\ref{ez2}) the gap amplitudes 
can be written as
\begin{eqnarray}
\label{irgapequ}
\Delta_{i} &=& - \: g_{i} \: \sum_{j \: \in \: {\cal R}} \: \Delta_{j} \: 
\sum_{\vec{k}} \: F(\vec{k}) \: \Psi_{i}(\vec{k}) \: \Psi_{j}(\vec{k}) \;
\end{eqnarray}       
for each irreducible representation {\em i}. In case of representations 
$B_{1}$ and $B_{2}$ the gap equation is scalar, 
while for representation $A_{1}$ we have to solve a coupled system 
consisting of three equations due to mixing of the corresponding three 
basis functions. A similar situation emerges for triplet pairing 
where two equations are coupled together since the corresponding irreducible 
representation of the states are two-dimensional. 

\begin{center}
\begin{table}
\caption{The basis functions $\Psi_{i}(\vec{k})$ and their notations,  
representations and the corresponding effective coupling constant $g_{i}$ 
entering the pair potential.}
\begin{tabular}{cccc}
${\cal R}$ & $\Psi_{i}(\vec{k})$ & notation & $g_{i}$ \\
\hline
$A_{1}$ & 1 & $s$ & $U_{0}$ \\
$A_{1}$ & $\frac{1}{2}(\cos{k_{x}}+\cos{k_{y}})$ & $s^{\ast}$ &
$4U_{1}$ \\
$A_{1}$ & $\cos{k_{x}}\!\cos{k_{y}}$ & $s_{xy}$ & $4U_{2}$ \\
$B_{1}$ & $\frac{1}{2}(\cos{k_{x}}-\cos{k_{y}})$ & $d_{x^{2}-y^{2}}$ &
$4U_{1}$ \\
$B_{2}$ & $\sin{k_{x}}\!\sin{k_{y}}$ & $d_{xy}$ & $4U_{2}$ \\
$E$ & $\frac{1}{2} \left( \begin{array}{c} \sin{k_{x}}+\sin{k_{y}} \\ 
\sin{k_{x}}-\sin{k_{y}} \end{array}
\right)$ & $\left( \begin{array}{c} p \\ \tilde{p} \end{array} \right)$ & 
$4U_{1}$\\
$E$ & $\frac{1}{2} \left( \begin{array}{c} \sin{k_{x}}\!\cos{k_{y}}+
\sin{k_{y}}\!\cos{k_{x}} \\ 
\sin{k_{x}}\!\cos{k_{y}}-\sin{k_{y}}\!\cos{k_{x}} \end{array}
\right)$ & $\left( \begin{array}{c} p_{2} \\ \tilde{p}_{2} \end{array} 
\right)$ & $8U_{2}$
\label{tablazat}
\end{tabular} 
\end{table}
\end{center}

The dispersion relation is given in terms of the real space hopping 
matrix elements as 
\begin{eqnarray}
\epsilon(\vec{k}) &=& t_{0} \: + \: \sum_{l} \: t_{l} 
\exp{\!(i\vec{k}\vec{R}_{i,i+l})} \;,
\label{ez3}
\end{eqnarray} 
where $\vec{R}_{i,i+l}$ denotes lattice vector pointing from site {\em i} to 
its $l^{th}$ neighbour ($l \!\geq\! 1$).

At a given temperature all the possible ordered phases can be found as the 
nontrivial solutions of Eqs.\ (\ref{opequation}) and (\ref{muequation}).
After the self-consistent solution, the stability of ordered phases of
different symmetry was carefully investigated. This has been performed
on the basis
of a comparative free-energy (or at $T \!=\! 0$ internal energy) analysis
(for the procedure see eg. (Dahm, Erdmenger, Scharnberg and Rieck, 1993)).
If the gap-equation presented more possible solutions connected to the same 
point of the phase diagram, the free-energy of each solution was determined 
and compared. For any solution, the free-energy relative to the normal
state per particle $\delta F$, can be derived by improving the
method of Gyorffy {\em et al.} (Gyorffy, Staunton and Stocks, 1991),
the final result being given by
\begin{eqnarray}
\label{freeenergy}
\delta F[\Delta(\vec{k}),\: \mu, \: T] &=& - \: \sum_{\vec{k}} \! \left\{ \:
 E(\vec{k}) \: - \: |\Omega(\vec{k})| \: \right\} \: + \: 2 \:
\sum_{\vec{k}} \: 
\left\{ \: \Delta^{2}(\vec{k}) \: F(\vec{k}) \: - \: \frac{1}{\beta} \: 
\ln{ \: \frac{1 \: + \: e^{- \: \beta \: E(\vec{k})}}{1 \: + \: e^{-
      \: \beta \:
|\Omega(\vec{k})|}}} \: \right\} \nonumber \\
&& \: + \: \sum_{\vec{k}} \: \sum_{\vec{q}} \: \Delta(\vec{q}) \:
F(\vec{q}) \: 
U(\vec{q}-\vec{k}) \: \Delta(\vec{k}) \: F(\vec{k}) \;.
\end{eqnarray}
In this manner, altering the interaction parameters, the superconducting 
phase diagram of the present model has been constructed systematically at 
different temperature and band-filling values and main superconducting
characteristics have been studied in every phase diagram domain in detail. 

\section{The obtained results}
\label{results}

\subsection{Numerical procedure}
\label{numproc}

To construct a phase diagram as presented above, first 
Eqs.\ (\ref{muequation}) and (\ref{irgapequ}) have been solved for the 
amplitudes $\Delta_{i}$ corresponding to different representations. Since the 
factor $F(\vec{k})$ occurring in these equations depends on all the 
$\Delta_{i}$ values, a self-consistent procedure has to be applied. This 
requirement has been fulfilled by the adoption of Broyden's 
algorithm (Numerical Recipies in Fortran, 1992), which assures the
global convergence. 

To start the algorithm, first a vector composed of $\Delta_{i}$ and $\mu$ 
denoted by $\underline{\Delta}^{(1)}_{i}$ is picked up arbitrarily at the 
fixed {\em T}, {\em n} and $g_{i}$ values. At this choice the possibility of 
mixing different symmetry representations is allowed with no restrictions. 
Then using this point as a zeroth guess for the solution of 
Eqs.\ (\ref{muequation}) and (\ref{irgapequ}) the nonlinear set of equations 
is solved numerically in order to get $\underline{\Delta}^{(1)}_{f}$. 
In the knowledge of $\underline{\Delta}^{(1)}_{f}$ the initial guess for the 
next iteration $\underline{\Delta}^{(2)}_{i}$ can be constructed in terms of  
the linear combination of $\underline{\Delta}^{(1)}_{i}$ and 
$\underline{\Delta}^{(1)}_{f}$. This linear combination provides the global 
nature of convergence (i.e. directional dependences are eliminated and the 
solution does not depend on the starting $\underline{\Delta}^{(1)}_{i}$ value).
Then, using $\underline{\Delta}^{(2)}_{i}$ as a starting point for the 
next iteration ($m\!=\!2$), the equations are solved again to obtain
$\underline{\Delta}^{(2)}_{f}$, in terms of which the starting point for the 
3{\em rd} iteration $\underline{\Delta}^{(3)}_{i}$ can be constructed, and so 
on. The above procedure is continued until self-consistency has been reached, 
that is, until at the end of the m{\em th} iteration the difference 
$\epsilon_{m}\!=\!|\underline{\Delta}^{(m)}_{f} \!-\! 
\underline{\Delta}^{(m)}_{i}|$ is less than a small, positive 
and {\it a priori} fixed value, $\epsilon$. In the present calculation 
$\epsilon$ is chosen to be $10^{-6}$. The required convergence is fulfilled 
in average after $m \sim 25-30$ iterations for the chosen value of $\epsilon$.

Because of the complexity of this analysis, only the possible superconducting 
phases were considered (i.e. the possible emergence of spin ({\em SDW}) and
charge ({\em CDW}) density wave type phases was discussed, although it was not 
investigated systematically in detail).

\subsection{Hamiltonian with nearest neighbour terms only}
\label{NNTB}

\begin{figure}[!ht]
\begin{center}
\epsfig{file=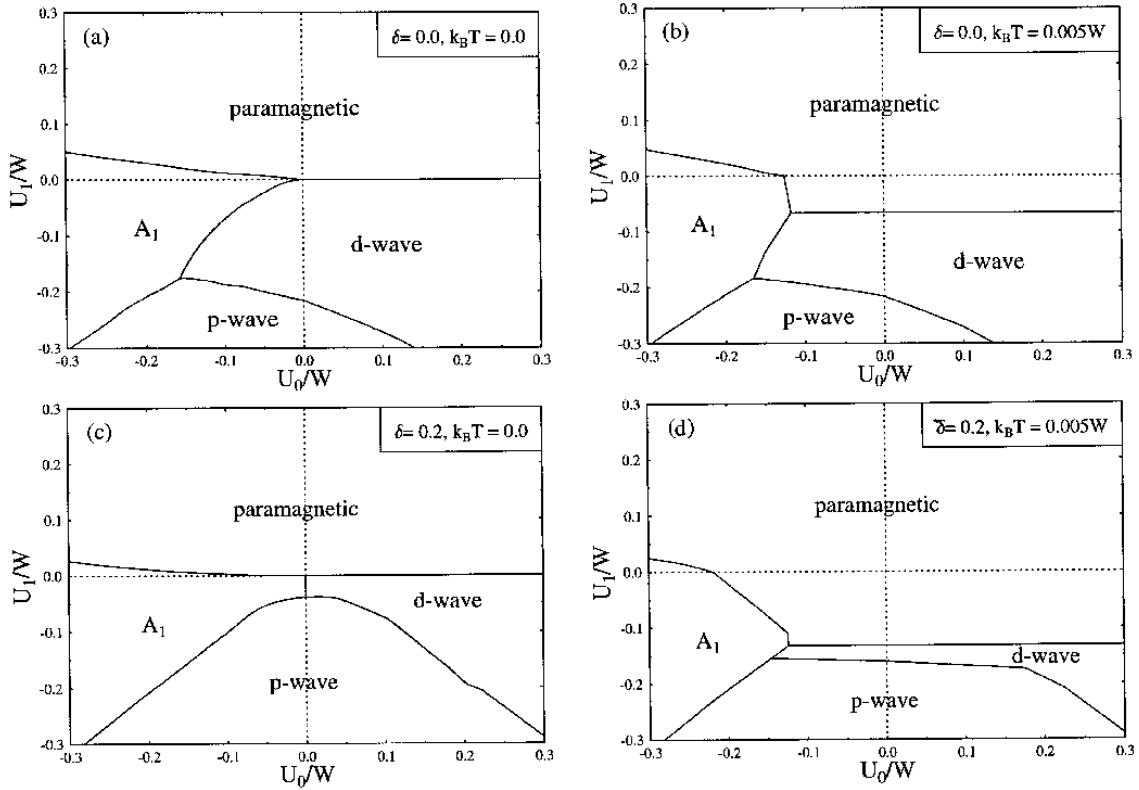, width=15cm}
\caption{Nearest-neighbour phase diagram at zero ($k_{B}T\!=\!0.0$) and 
nonzero ($k_{B}T\!=\!0.005W$) temperatures for different dopings 
($\delta$), using dispersion containing only nearest neighbour hopping 
terms. Dotted lines mean coordinate axis, solid lines represent 
boundaries between different symmetry phases. For notations 
$A_{1}$,~{\em d},~{\em p}, and {\em paramagnetic} see the text. }
\label{fig1}
\end{center}
\end{figure}

We started our study with the analysis of pure {\em NN}-case where 
$U_2\!=\!0$ is satisfied and non-vanishing 
hopping matrix elements were taken into consideration only between 
nearest neighbouring sites, that is, $t_{l} \!\ne\! 0$ for $l\!=\!1$ and 
$t_{l} \!=\! 0$ otherwise. In what follows, all the couplings are normalized 
by the bandwidth, $W \!=\! 8t_{1}$ (for example, in the case of 
{\em Bi2212} we 
have $W \!=\! 1.192$ eV extracted from {\em ARPES} data). The obtained results 
are plotted in Fig.\ \ref{fig1}. Here four separate plots are presented at 
zero (Fig.\ \ref{fig1}(a) and (c)) and nonzero (Fig.\ \ref{fig1}(b) and (d)) 
temperatures for half filling ($\delta \!=\! 0.0$, Fig.\ \ref{fig1}(a),(b)) 
and away from half filling ($\delta \!\ne\! 0.0$, Fig.\ \ref{fig1}(c),(d)). 

Concerning the notations, {\it paramagnetic} refers to the normal state of 
the model, {\it d-wave} and {\it p-wave} stand for order parameters with
$(\cos{k_{x}}\!-\!\cos{k_{y}})$ and $(\sin{k_{x}} \!+\! \sin{k_{y}})$ 
{\bf k}-dependence, respectively, while $A_{1}$ denoting the $A_{1}$-symmetry 
solution of the gap equations represents in this paragraph 
(in Fig.\ \ref{fig1}) a standard {\em s}-wave component for plots (a)
and (b), and an $(s,s^{\ast})$ admixture for plots (c) and (d). Here 
$s^{\ast}$ stands for the extended {\em s}-wave solution. 

In Fig.\ \ref{fig1}(a) the $T\!=\!0$ ground state phase diagram at 
half-filling is shown. As it can be seen, if either the on-site $U_{0}$ or 
the intersite $U_{1}$ interaction is attractive superconductivity emerges, 
where the symmetry of the stable, ordered phase revealed by the $\delta F$ 
analysis is $A_{1}$- or $d_{x^{2}-y^{2}}$ for resonable values of $|U_1/W|$. 
Furthermore, if both interaction parameters are attractive a competition 
between different symmetry pairing states is observed. It can also be
seen that increasing the value of $|U_{0}|$, triplet pairing is no
longer favored. This means particularly, that repulsive on-site
interaction stabilizes the superconducting state of $d_{x^{2}-y^{2}}$-symmetry.
 
Comparing these facts with the preliminary results of Micnas 
{\em et al.}, a qualitative agreement can be observed. Here we 
should mention that taking spin and charge density wave ordering possibilities 
also into account, the paramagnetic region of Fig.\ \ref{fig1}(a) can be 
filled up by these phases not affecting the main superconducting regions of 
the plot. However, the occurrence of pure $d_{x^{2}-y^{2}}$ state
instead of an {\em s-d} mixture (Micnas, Ranninger, Robaszkiewicz and
Tabor, 1988, Micnas, J. Ranninger and S. Robaszkiewicz, 1989) is found
in Fig.\ \ref{fig1}.{\em a}, 
which is in accordance with the results of Fehrenbacher and Norman 
(Fehrenbacher {\em et al.}), O'Donovan and Carbotte (O'Donovan {\em et al.}) 
and also with group-theoretical studies (Wenger and \"Ostlund, 1993).

With the non half-filled band case at zero temperature 
(Fig.\ \ref{fig1}(c) at $\delta\!=\!0.2$), we start to explore a
parameter domain of the phase diagram which has not been investigated 
{\em systematically} so far in the literature. As it can be seen from
the plot, the symmetry of the pairing state is strongly doping dependent;
the $A_{1}$- and $p$-wave solutions, with increasing $\delta$, are gradually 
overwhelming the superconducting phase of $d_{x^{2}-y^{2}}$ symmetry, as
it is predicted by Micnas {\em et al.}. On the other hand, moving away from
half-filling an $s^{\ast}$ component of the $A_{1}$-symmetry solution
sets in, suggesting that the rigorous constraints on {\em s}-wave pairing 
deduced by Zhang for the standard Hubbard model with on-site interaction 
only (Zhang, 1990), is also applicable in the extended Hubbard model case.

The $T \!\ne\! 0$ counterparts of Fig.\ \ref{fig1}(a) and (c) are  
Fig.\ \ref{fig1}(b) and (d), respectively, keeping the same notations for
the labeling of superconducting phases. In Fig.\ \ref{fig1}(b) the phase 
diagram for the half-filled band is depicted. It can be observed, 
around $|U_{0}/W| \!\approx\! 0.15$, that with increasing temperature  
the stable $A_{1}$- and $d_{x^{2}-y^{2}}$-symmetry superconducting
phases are shifted and favored over the $p$-wave solution. This fact leads 
to an interesting feature, namely, at fixed 
band-filling and coupling constants, decreasing the temperature from  
$T_{c}^{(1)}$ defined as the critical temperature for a paramagnet - 
superconductor type transition, an additional second phase transition of 
superconductor - superconductor type appears at a lower temperature 
$T_{c}^{(2)} \!<\! T_{c}^{(1)}$. This is a first-order transition between two 
superconducting phases of different symmetry close to the phase boundary 
lines. A similar double transition between {\em s} - and 
$d_{x^{2}-y^{2}}$-wave states has been deduced by Dahm {\em et al.}
in case of superconducting states emerging from spin-fluctuation mechanism. 

Away from half-filling (Fig.\ \ref{fig1}(d)), the increasing value
of  doping modifies the $p$-wave region of the phase diagram, however, 
the rate of paramagnetic phase emergence as the function of {\em T} is
strongly doping dependent. Furthermore, at a fixed value of the
doping, for strong 
repulsive on-site interactions and not excessively low $U_{1} \!<\! 0$ values,
the $d_{x^2-y^2}$-symmetry phase represents not only a possible solution but
it is the most stable superconducting phase among all the solutions of the gap
equation. Qualitatively similar behaviour has been conjectured in various 
circumstances (Mierzejewski and Zielinski, 1995, Gufan, Vereshkov,
Toledano, Mettout, Bouzerar and Lorman, 1995).
Concerning the spin and charge density wave states not taken explicitly into 
account in Fig.\ \ref{fig1}, we expect their influence to be minor in the
non-paramagnetic domain of the phase diagram. In this respect we would like to 
note that the nesting condition is hindered by doping in general, so its 
effect is reduced at low temperatures (Inaba, Matsukawa, Saitoh and
Fukuyama, 1996) and following this line, doping usually enhances the 
emergence of superconductivity (Scalapino, 1995, Iglesias, Bernhard
and Gusmao, 1995).

\subsection{Nearest-neighbour interactions with more than nearest neighbour 
hopping terms}
\label{NNTBF}

From this paragraph on, besides {\em NN} terms, {\em NNN}
contributions are also incorporated in the model Hamiltonian. As a
first step, in order to obtain a
clear image about modifications arising in the phase diagrams {\it purely} due
to the presence of hopping terms exceeding {\em NN} distance in range, all the
previously used interaction terms are kept unchanged but a more complex 
kinetic energy contribution is introduced in the present paragraph.
This step also enables us to get an overall picture on how properties related 
to superconductivity are modified by DOS effects due to 
long-range hopping, without mixing these effects with the effects of {\em NNN} 
two-particle interaction terms. To accentuate the relevance to high-$T_{c}$ 
materials, we considered a dispersion relation obtained as a tight-binding fit 
to normal state {\em ARPES} data taken on {\em Bi2212} with nonzero hopping 
elements up to the fifth neighbours. The numerical values of the used $t_{l}$ 
terms can be found in Ref.\  (Fehrenbacher and Norman, 1995). Here we
note, that this new, extended 
dispersion breaks the $\delta \!\rightarrow\! -\delta$ symmetry of the phase 
diagrams and in what follows, only hole-doped cases are discussed. 

\begin{figure}[!ht]
\begin{center}
\epsfig{file=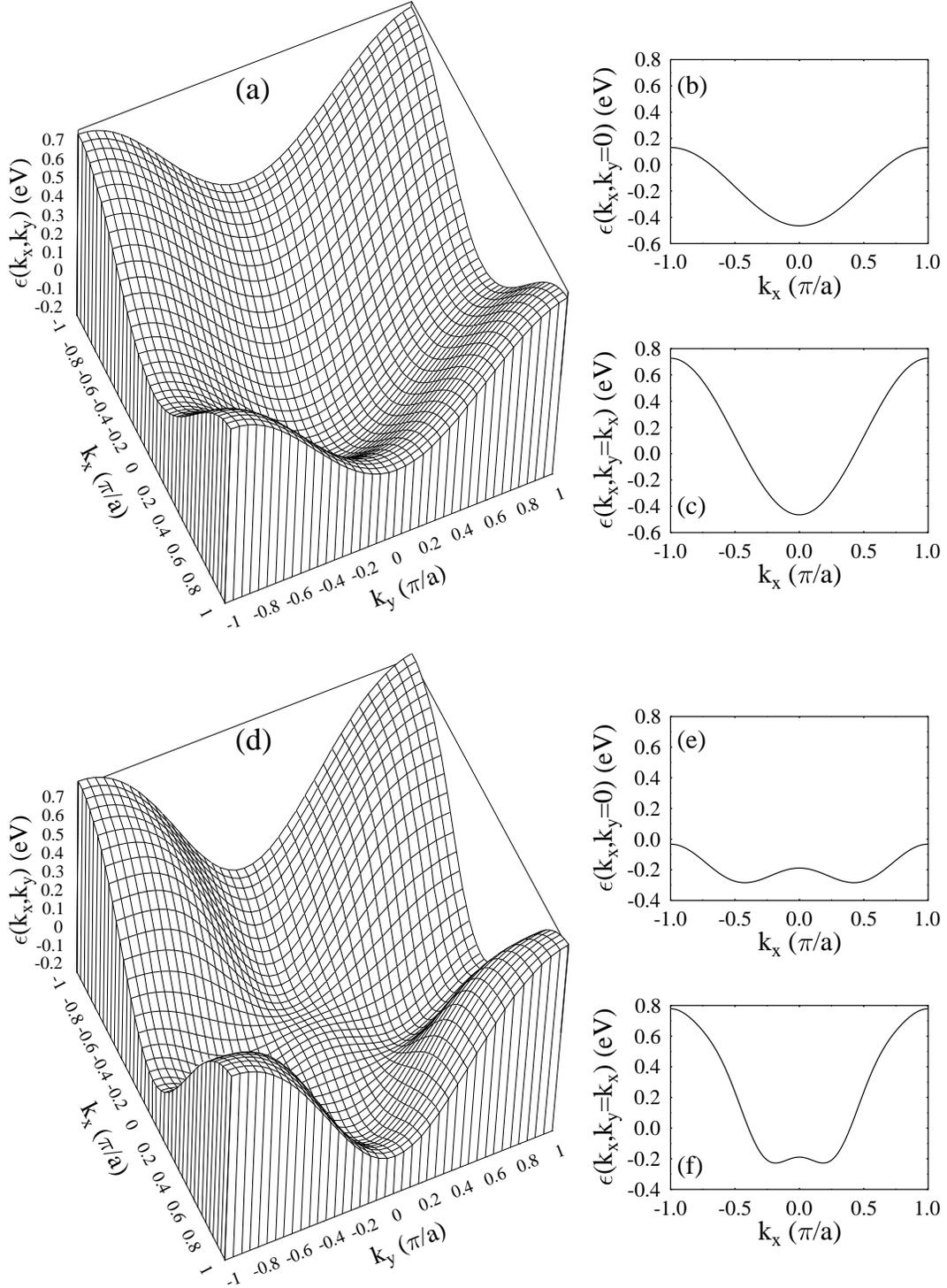, angle=180, width=15cm, bbllx=22, bblly=40,
  bburx=560, bbury=740}
\caption{Quaziparticle dispersion with nearest neighbour hopping terms 
[plot (a)] and beyond nearest neighbour hopping terms [plot (d)] in 
the first Brillouin zone. The insets show the dispersion in special 
directions, plots (b) and (e) are taken in the $(\pi,0)$ while plots 
(c) and (f) in the $(\pi,\pi)$ directions. }
\label{fig2}
\end{center}
\end{figure}

To show the source of modifications due to the extension in the hopping range,
we compared in Fig.\ \ref{fig2} the dispersion containing only {\em NN}, and 
the dispersion containing higher order contributions as well. The insets 
present energy dispersions along different directions of the first Brillouin
zone. At first glance hopping terms exceeding {\em NN} distance in range seem
to represent negligibly small corrections (for example in the present 
calculation $|t_2|/|t_1| \! \sim \! 0.25$ and $|t_3|/|t_1| \! \sim \! 0.08$).
This is not true, however, in the process of calculating thermodynamic
averages, since their effects during {\bf k}-integrations are extraordinarily
enhanced by the supplementary extrema introduced to the dispersion. As it 
can be seen, the characteristic change occurs at small
{\bf k}'s in the vicinity of the origin, with the flattening of the 
dispersion in this region. This leads to the fact (comparing 
Fig.\ \ref{fig2}(e) and (f) with Fig.\ \ref{fig2}(b) and (c)) that the region 
where ${\cal N}^{-1} \!\sim\! \partial \epsilon (\vec{k})/ \partial \vec{k} 
\!\simeq\! 0 $ holds, is much larger for the $t_{l} \!\ne\! 0$ 
$(l \!\leq\! 5)$ 
case than in the $t_{l} \!\ne\! 0$ $(l \!\leq\! 1)$ case. Note, that all the 
integrands within the expressions that must be evaluated are proportional to 
${\cal N}$. As a consequence, main DOS effect contributions like
van  Hove singularities (Blumberg {\em et al.}), saddle points
(Abrikosov, 1995) and peaks (Cappelluti and Pietronero, 1996) are 
strongly wiped out by taking $t_{l} \!=\! 0$ for $l \!\geq\! 2$, a choice that 
could lead to completely misleading conclusions in connection with 
the studied materials. 

\begin{figure}[!ht]
\begin{center}
\epsfig{file=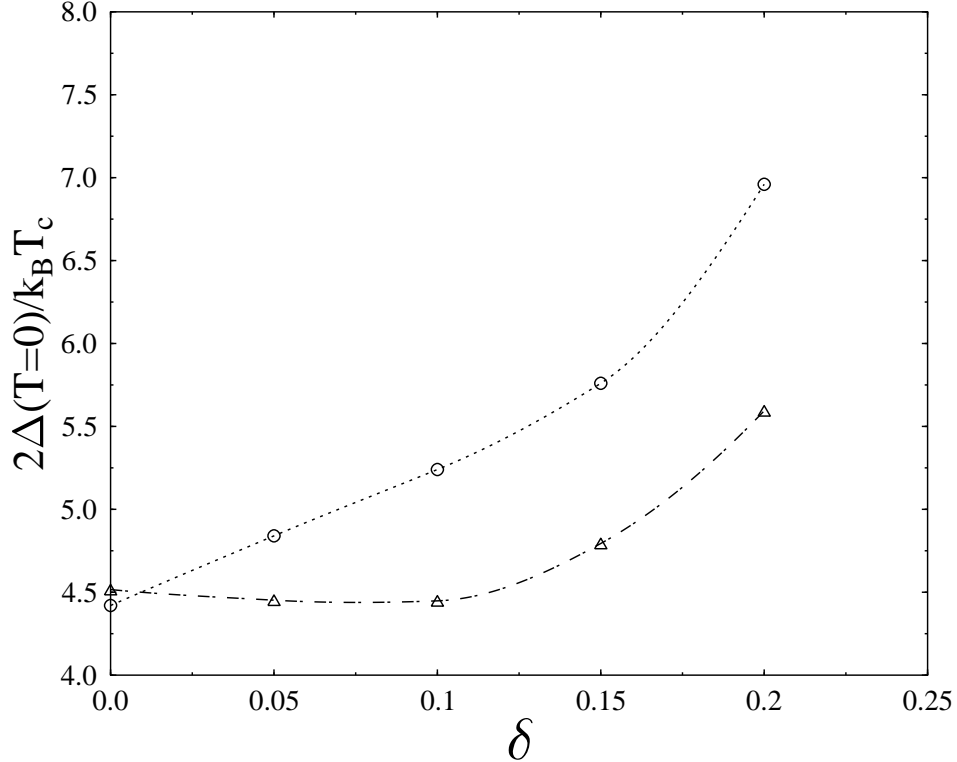, angle=270, width=13cm, bbllx=200, bblly=85,
  bburx=510, bbury=465}
\caption{Doping dependence of $2\Delta(T\!=\!0)/k_{B}T_{c}$ for the
stable $d_{x^{2}-y^{2}}$ gap at a fixed set of coupling constants
$U_{0} \!=\! 0.2W$, $U_{1} \!=\! -0.15W$ and $U_{2}\!=\!0.0$. The
calculated values are denoted by triangles and circles for the cases 
where only nearest neighbour and further than nearest neighbour hopping 
terms were taken into consideration, respectively.}
\label{fig3}
\end{center}
\end{figure}

To illustrate the effect of dispersion modifications presented in
Fig.\ \ref{fig2}. on different quantities of interest, the ratio 
$R\!=\!2\Delta_{0}/k_{B}T_{c}$, ($\Delta_{0}$ is the zero temperature gap
ampitude) versus doping is plotted in Fig.\ \ref{fig3}. for a stable 
$d_{x^{2}-y^{2}}$-symmetry
superconducting state of the phase diagram. In this plot open triangles are
calculated with the dispersion used in Sec.\ \ref{NNTB}, while open circles
are obtained applying the extended dispersion used in the present Section. The
effect of hopping terms exceeding $NN$ distance in range can easily be 
seen. Besides the clear absolute value differences we note that the curve 
connecting the open triangles has a minimum at about $\delta\!=\!0.1$,
while the other line takes its least value at half filling ($\delta\!=\!0.0$).
Here we also would like to underline that the $R \!\approx\! 5.0$ value in the 
optimally doped regime ($\delta \!\approx\! 0.12$) qualitatively agrees 
with predictions of other authors (Abrikosov, 1995, Dagotto, Nazarenko
and Moreo, 1995).
Furthermore, Fig.\ \ref{fig3}. nicely illustrates that higher
amount of doping leads to a greater value of {\em R}, an enhancement that is 
clearly increased by the $t_{l} \!\ne\! 0$ ($l \!\geq\! 2$) terms.

\begin{figure}[!ht]
\begin{center}
\epsfig{file=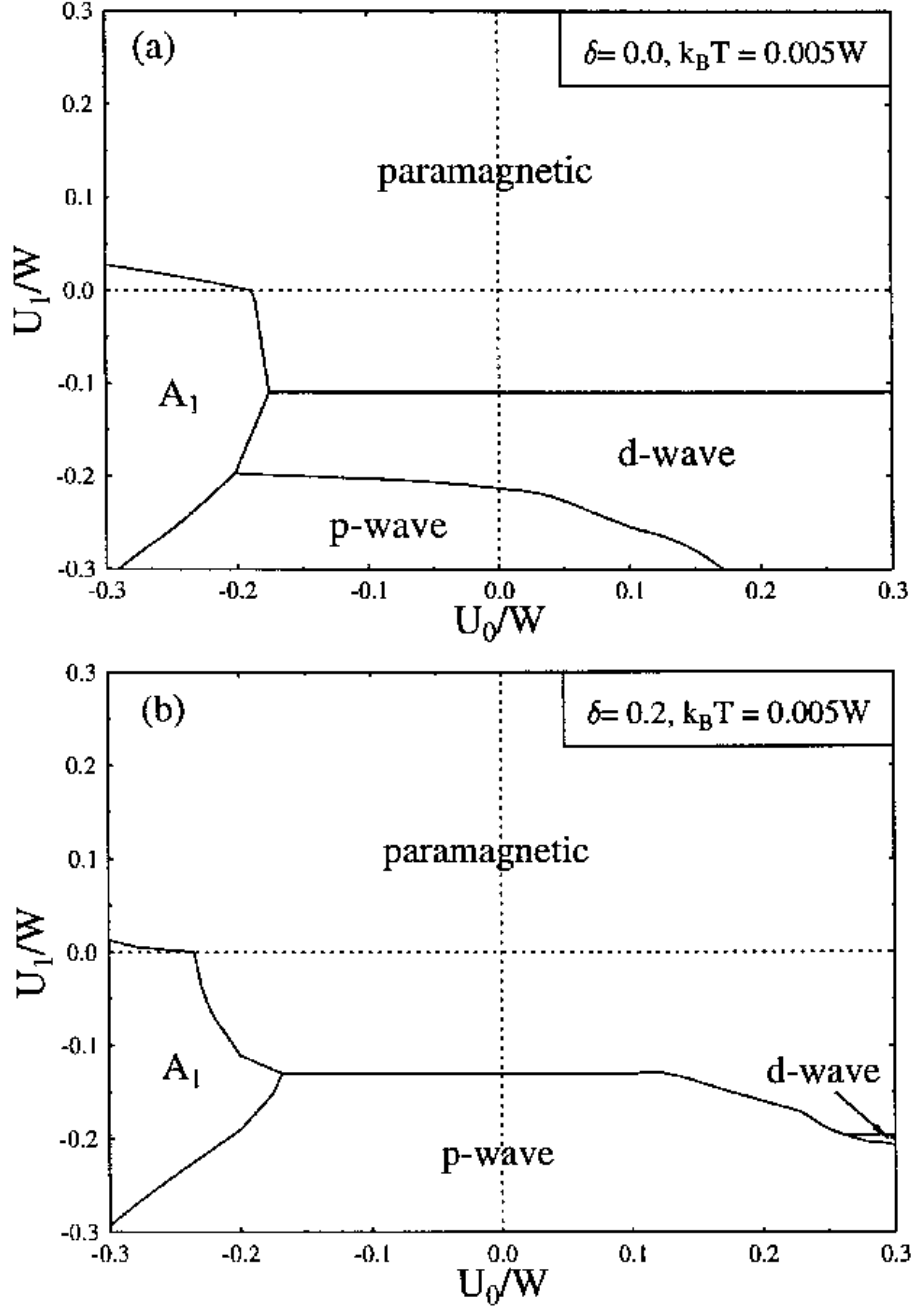, width=13cm, bbllx=170, bblly=210,
  bburx=440, bbury=580}
\caption{Phase diagram at $k_{B}T\!=\!0.005W$ temperature for different 
dopings, using dispersion containing hopping terms beyond nearest 
neighbouring ones. Notations and curves have the same meaning as in 
Fig.\ \protect\ref{fig1}. }
\label{fig4}
\end{center}
\end{figure}

Concerning the modifications produced in the phase diagram, the obtained 
results are exemplified at nonzero temperature ($k_{B}T/W \!=\! 0.005$) with 
Fig.\ \ref{fig4}(a) and (b) for $\delta \!=\! 0.0$ and $\delta \!=\! 0.2$, 
respectively, which should be compared to Fig.\ \ref{fig1}(b) and (d) 
plotted at the same values of {\em T} and $\delta$. As it can be seen, the 
effects become more robust with increasing the doping. The extension of the
$A_1$-symmetry domain becomes reduced and the $d_{x^2-y^2}$ phase boundary is
withdrawn more accentuately with increasing temperature in the presence of 
{\em NNN} hoppings.

All these theoretical results are perfectly compatible with and
underline aspects
observed in an increasing number of experimental data covering a large 
spectrum and suggesting the importance of {\em NNN} hopping contributions in 
building up superconducting properties: inelastic neutron 
scattering (Bourges, Regnault, Sidis and Vettier, 1996),
magneto-resistance and conductivity (Mashimoto, Nakao, Kado and 
Koshizuka, 1996), thermopower, resistivity (Gasumyants, Ageev,
Vladimirskaya, Smirnov, Kazanskiy and Kaydanov, 1996), and Hall 
effect (Hopfengartner, Leghissa, Kreiselmeyer, Holzapfel,
Schmitt and Ischenko, 1993) measurements.
Theoretical support of this line (Cappelluti {\em et al.}, Normand,
H. Kohno and H. Fukuyama, 1996) is also present (see also the Introduction).

\subsection{Influence of next-nearest neighbour interactions}
\label{NNNTBF}
         
\begin{figure}[!ht]
\begin{center}
\epsfig{file=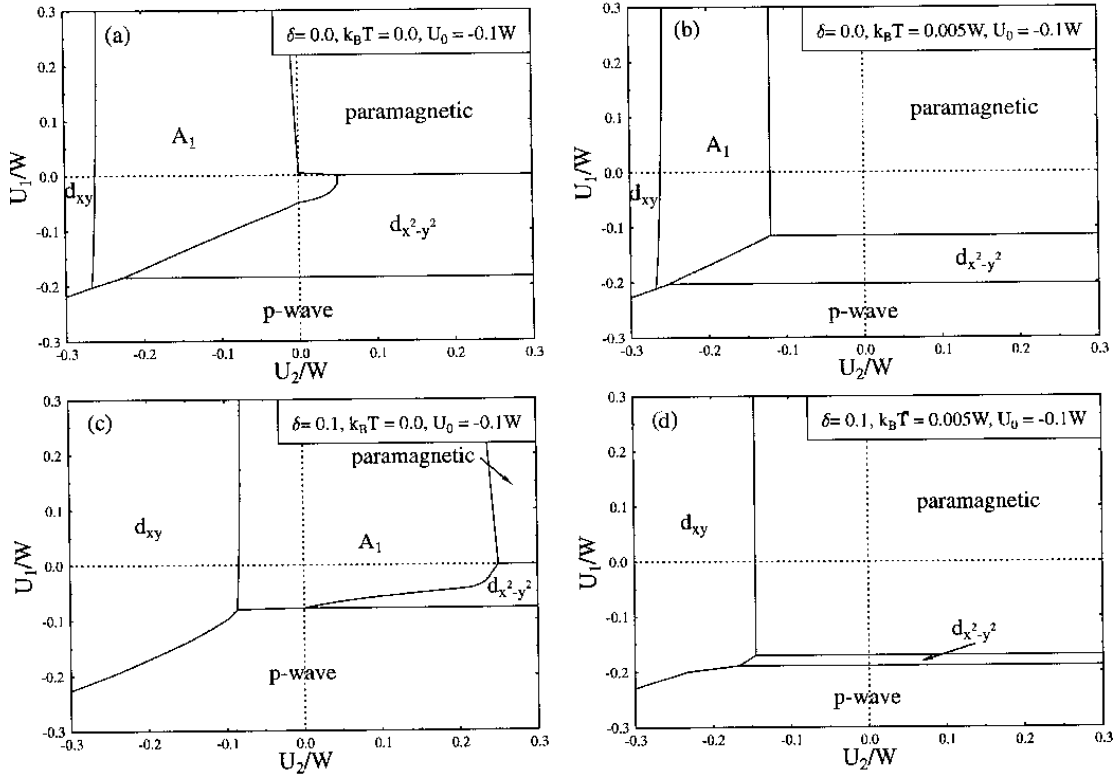, width=15cm, bbllx=60, bblly=225,
  bburx=540, bbury=570}
\caption{Phase diagram with next-nearest neighbour couplings 
at zero ($k_{B}T\!=\!0.0$) and nonzero ($k_{B}T\!=\!0.005W$) temperatures 
for different dopings and fixed on-site attraction, $U_{0}\!=\! -0.1W$, 
for a dispersion containing also hopping terms beyond nearest neighbouring 
ones. Dotted lines mean coordinate axis and solid lines represent 
boundaries between different symmetry phases. The meaning of notations 
$d_{x^{2}-y^{2}}$, $d_{xy}$, $A_{1}$, {\em p-wave} and {\em paramagnetic} 
is the same as in the text.}
\label{fig5}
\end{center}
\end{figure}

\begin{figure}[!ht]
\begin{center}
\epsfig{file=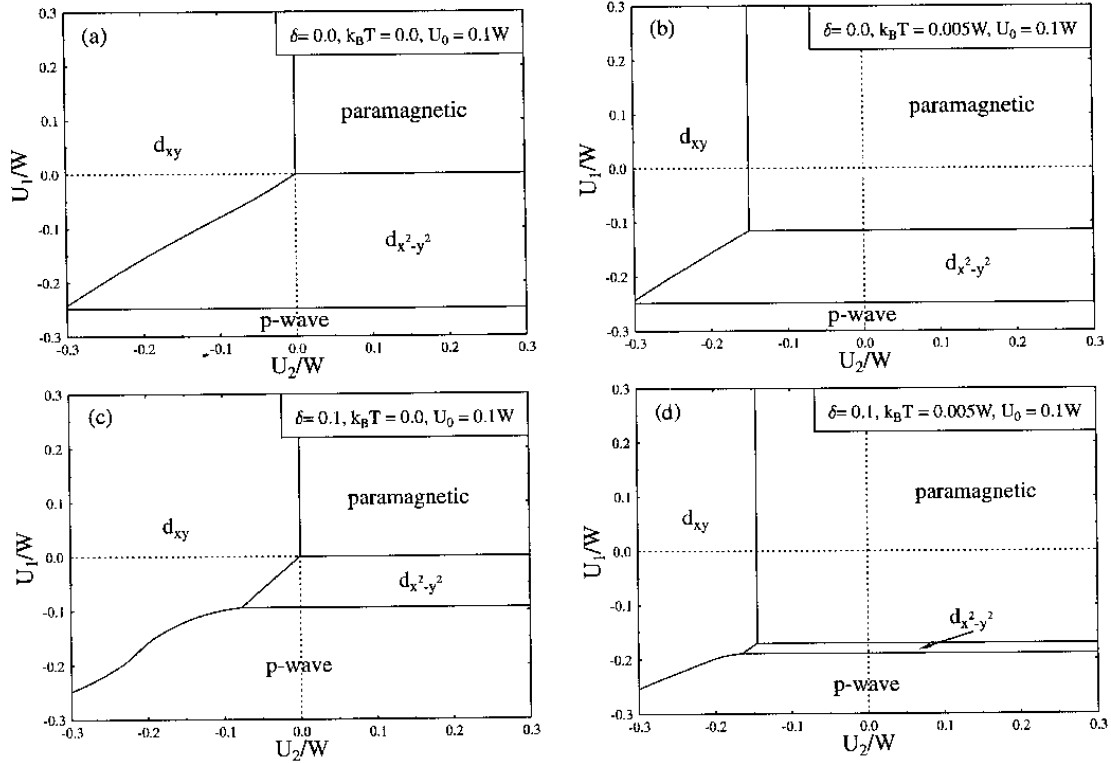, width=15cm, bbllx=60, bblly=225,
  bburx=540, bbury=570}
\caption{Phase diagram with next-nearest neighbour couplings 
at zero ($k_{B}T\!=\!0.0$) and nonzero ($k_{B}T\!=\!0.005W$) temperatures 
for different dopings and fixed on-site repulsion, $U_{0}\!=\! 0.1W$, for 
a dispersion containing also hopping terms beyond nearest neighbouring ones. 
Notations and curves have the same meaning as in Fig.\ \protect\ref{fig5}.}
\label{fig6}
\end{center}
\end{figure}

From now on, the attention is focused on the changes in  
superconducting properties introduced by the nonzero value of {\em NNN}  
interactions. The dispersion used in Sec. \ref{NNTBF} has been kept 
unchanged for the present Section and {\em NNN} coupling with 
$U_{2} \!\ne\! 0$ has been considered. The obtained 
results regarding the superconducting phases at different temperature
and doping values are shown in Fig.\ \ref{fig5} and Fig.\ \ref{fig6} for
attractive and repulsive on-site interactions, respectively. Concerning the
notations we would like to underline the following aspects. 
The presence of nonzero $U_{2}$ term results in new singlet
superconducting states characterized by {\bf k}-dependent order parameters
proportional to $\sin{k_{x}}\!\sin{k_{y}}$ or $\cos{k_{x}}\!\cos{k_{y}}$ 
denoted by $d_{xy}$ or $s_{xy}$, respectively, and new triplet
superconducting states with ($\sin{k_{x}}\!\cos{k_{y}} \pm 
\sin{k_{y}}\!\cos{k_{x}}$) type of {\bf k}-dependence. For the last two 
states the $p_{2}$- and $\tilde{p}_{2}$-wave notations are used. The 
symmetry classification of the new states can be found in Table \ref{tablazat}.
In what follows, considering the phase diagrams, the notation $d_{xy}$ stands 
for an order parameter having $B_{2}$ symmetry, $d_{x^{2}-y^{2}}$ means an 
order parameter of $B_{1}$ symmetry and the symbol $A_{1}$ refers to an order 
parameter having $s$,$s^{\ast}$ and $s_{xy}$ components. Further 
notations are the same as in Fig.\ \ref{fig1}. In the presentation, for the 
sake of clarity, we concentrate on fixed values of $U_{0}$ with opposite signs 
($U_0/W \!=\!-0.1$ and $U_0/W \!=\! 0.1$ in Fig.\ \ref{fig5} and 
Fig.\ \ref{fig6}, respectively), in order to maintain the two-dimensional 
structure of figures. 

The $T\!=\!0$ ground state phase diagram for different dopings ($\delta \!=\!
0.0$ and $\delta \!=\! 0.1$) in the presence of attractive on-site interaction
presented in Fig.\ \ref{fig5}(a) and (c) shows a richness of different 
superconducting symmetry species. At this point we would like to emphasize 
the stability of the $A_{1}$-symmetry pairing state containing a
coexistence of $s^{\ast}$, $s_{xy}$ and {\em s}-wave components. It is 
interesting to note that the $s^{\ast}$ and $s_{xy}$ components are governed 
by {\em NN} and {\em NNN} interaction, respectively, while the conventional 
{\em s}-wave part is primarily determined by the on-site interaction.
As a consequence, despite the repulsive intersite couplings, the
$A_{1}$-symmetry superconducting phase stabilizes due to the compensation of 
repulsions by the on-site attraction. In this compensation doping is of great 
importance, since a small increase in its value induces the emergence of
$A_1$-symmetry pairing state in the non-superconducting region of the
phase diagram, as shown in Fig.\ \ref{fig5}(c). 

The structure of the phase diagram at a higher, nonzero temperature is 
exemplified in Fig.\ \ref{fig5}(b) and (d). While the $A_{1}$ phase is
rapidly destroyed by increasing the temperature, the $d_{xy}$ phase entirely 
induced by {\em NNN} terms shows a more accentuated resistance against 
thermal pair-breaking effects. In this way, for large values of $|U_{2}|/W$ 
the stability of the $d_{xy}$-symmetry  pairing state can be coherently
deduced within the presented model, a fact that underlines the correctness
of the conjecture made by Fehrenbacher and Norman (Fehrenbacher {\em
  et al.} related to the
stability of the $d_{xy}$ state.

Concerning triplet pairing we note that the states with $\tilde{p}$-, 
$p_{2}$- and $\tilde{p}_{2}$-symmetry never have lower free-energy in the 
coupling constants' space under study than the triplet pairing state of 
standard $p$-wave symmetry.
 
If now the $U_{0}\!>\!0.0$ case is considered, the phase diagram significantly 
changes, as it is shown in Fig.\ \ref{fig6}. The most important feature of 
this figure is, that the $A_{1}$-symmetry solution never reaches the
level of a 
stable superconducting state within the analyzed $U_{2}$ region for 
$|U_1|\!\ll\!1.0$ for slight dopings. This result is in accordance with the 
observation of Fehrenbacher {\em et al.}, where it is claimed that in
the presence of 
on-site repulsion with small {\em NNN} attractions ($|U_2|/W\!\ll\!1.0$), the 
stable superconducting state of $A_{1}$-symmetry can be realized in the limit 
of $|U_{1}| \!\rightarrow\! 0.0$ only at relatively high values of the doping.
 
Figure\ \ref{fig6}(a) and (c) show the $T \!=\! 0$ phase diagram of
the model with $U_{0} \!=\! 0.1W$ at two doping values ($\delta \!=\! 0.0$ and
$\delta \!=\! 0.1$) in the $(U_{1}/W, U_{2}/W)$ plane. The phase boundary 
between the $d_{x^{2}-y^{2}}$ and $d_{xy}$ phases moves slowly into the 
domain of $d_{x^{2}-y^{2}}$ phase with increasing values of the doping, 
that is, higher doping favors states of $d_{xy}$-symmetry over states of 
$d_{x^{2}-y^{2}}$-symmetry. Furthermore, as it can be observed in 
Fig.\ \ref{fig6}(a) and (c), the $d_{x^2-y^2}$-$d_{xy}$ phase boundary is 
linear and strongly affected by doping. For small values 
of $\delta$ it coincides with the $U_{1} \!=\!\alpha U_{2}$ line with 
$\alpha \simeq 1.0$, while for larger doping values it is shifted away and 
$\alpha$ becomes smaller than $1.0$. This leads to the following interesting 
feature: changing the doping continuously the symmetry of the stable 
superconducting state changes and a transition between the two different 
{\em d}-like pairing states (from $B_{2}$-symmetry to $B_{1}$-symmetry) takes 
place for small, fixed values of the coupling constants. This transition line
within the $T\!=\!0$ phase diagram, suggested by detailed
investigations, is of first order. 

Examining phase boundaries between the $p$ and $d_{xy}$ or 
$d_{x^{2}-y^{2}}$ superconducting states, one can see them moving upward with 
increasing doping into the $d_{x^{2}-y^{2}}$-symmetry phase. However, large 
{\em NNN} interaction ($|U_{2}|/W\!\gg\!1$) and small {\em NN} attraction 
($|U_{1}|/W\!\ll\!1$) favors superconducting pairing state of 
$d_{x^{2}-y^{2}}$ or $d_{xy}$ symmetry, depending on the sign of $U_{2}$,
an observation that holds even for heavily doped systems.   

Turning now to the $T \!\ne\! 0$ case depicted in Fig.\ \ref{fig6}(b) and (d), 
it can be observed that the triple-point moves downward in the 
$U_{1},U_{2}\!<\!0.0$ region with increasing the temperature. The phase 
boundary of the $d_{x^2-y^2}$ phase modifies more rapidly with 
increasing {\em T}, suggesting that in the large $|U_{2}|$ region (with 
$U_{2}/W \!<\! 0.0$ and $|U_{1}|/W \!\ll\! 1.0$) the highest critical 
temperature can be reached with a $d_{xy}$ phase. The stability of the 
$d_{xy}$ phase against thermal fluctuations, in comparison with the same 
stability of $d_{x^{2}-y^{2}}$ phase, is again striking. 

In the light of Figs.\ \ref{fig5} and \ref{fig6}, one can conclude 
that small $|U_{1}|/W$ and large attractive {\em NNN} coupling 
$U_{2}$ favors $d_{xy}$ pairing, while small $|U_{2}|/W$ and 
strong attractive {\em NN} coupling $U_{1}$ gives rise to stable 
$d_{x^{2}-y^{2}}$ pairing. These results emphasize the correctness of the
conjecture made by Wenger and \"Ostlund (Wenger {\em et al.}) related
to the stability of different {\em d} symmetry species, who also
claimed that third-neighbour attractive interaction could lead to a
stable {\bf k}-dependent {\em s}-wave within the phase diagram,
especially in the presence of repulsive closer-neighbour interactions.

\begin{figure}[!ht]
\begin{center}
\epsfig{file=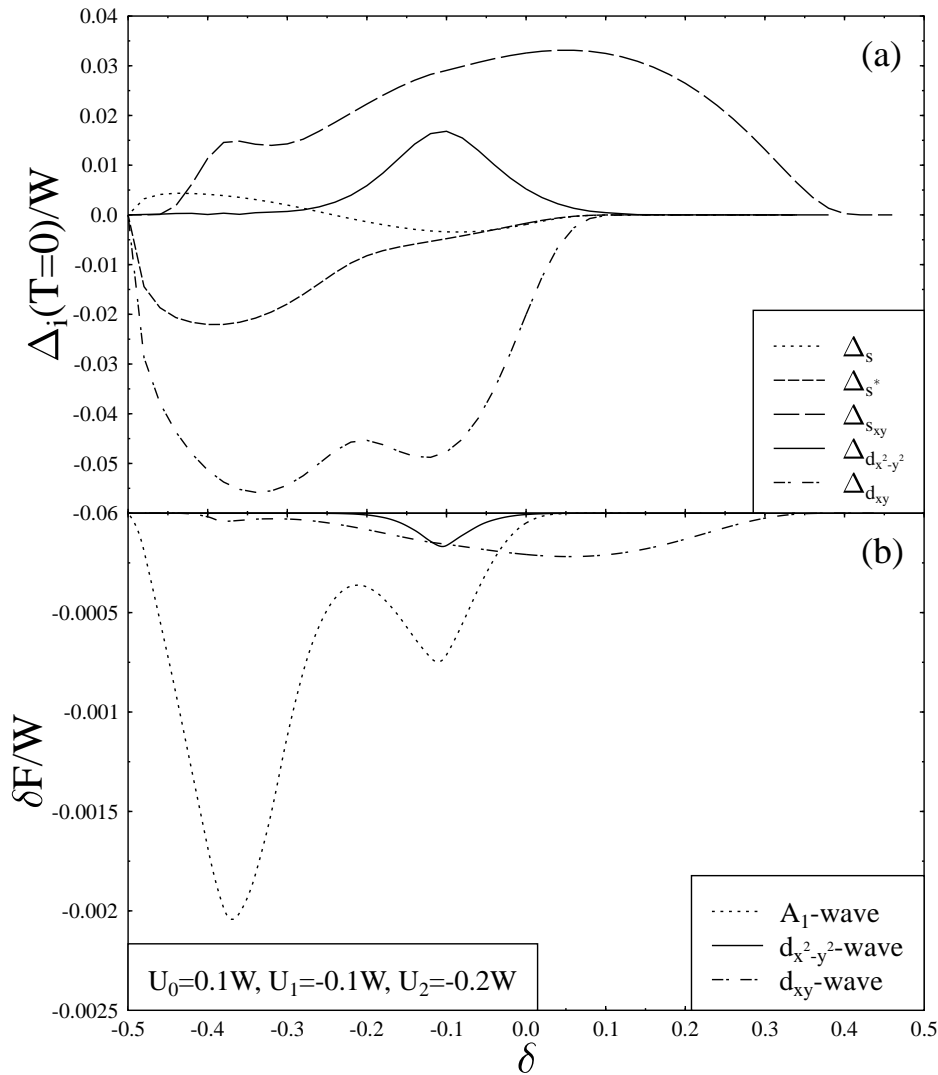, width=13cm, angle=180, bbllx=140, bblly=150,
  bburx=530, bbury=590}
\caption{Doping dependence of gap amplitudes [plot (a)] and of 
corresponding free-energies [plot (b)] at a fixed set of coupling 
constants $U_{0}\!=\! 0.1W$, $U_{1}\!=\!-0.1W$, $U_{2}\!=\!-0.2W$ for the 
dispersion containing hopping terms beyond nearest neighbouring ones.}
\label{fig7}
\end{center}
\end{figure}

The electron doping versus hole doping behaviour is examplified with 
Fig.\ \ref{fig7}, where doping dependence of gap amplitudes and free-energy 
are shown for a certain set of coupling constants. Having a glance at  
Fig.\ \ref{fig7}(a) one can see that besides the gap amplitudes,
phase also changes (i.e. sign change of $\Delta_{i}$ in the presented
figure) emerge 
for different solutions at a critical value of doping. More interestingly, 
Fig.\ \ref{fig7}(b) predicts the stable superconducting state to have 
$A_{1}$-symmetry in the electron doped regime even for repulsive on-site 
couplings. This suggests that the symmetry of the pairing state in electron 
doped materials (for example in {\em NCCO}) might be explained without 
assuming attracion in the on-site interaction channel. Systematic study of 
the electron doped case is under progress.  

\begin{figure}[!ht]
\begin{center}
\epsfig{file=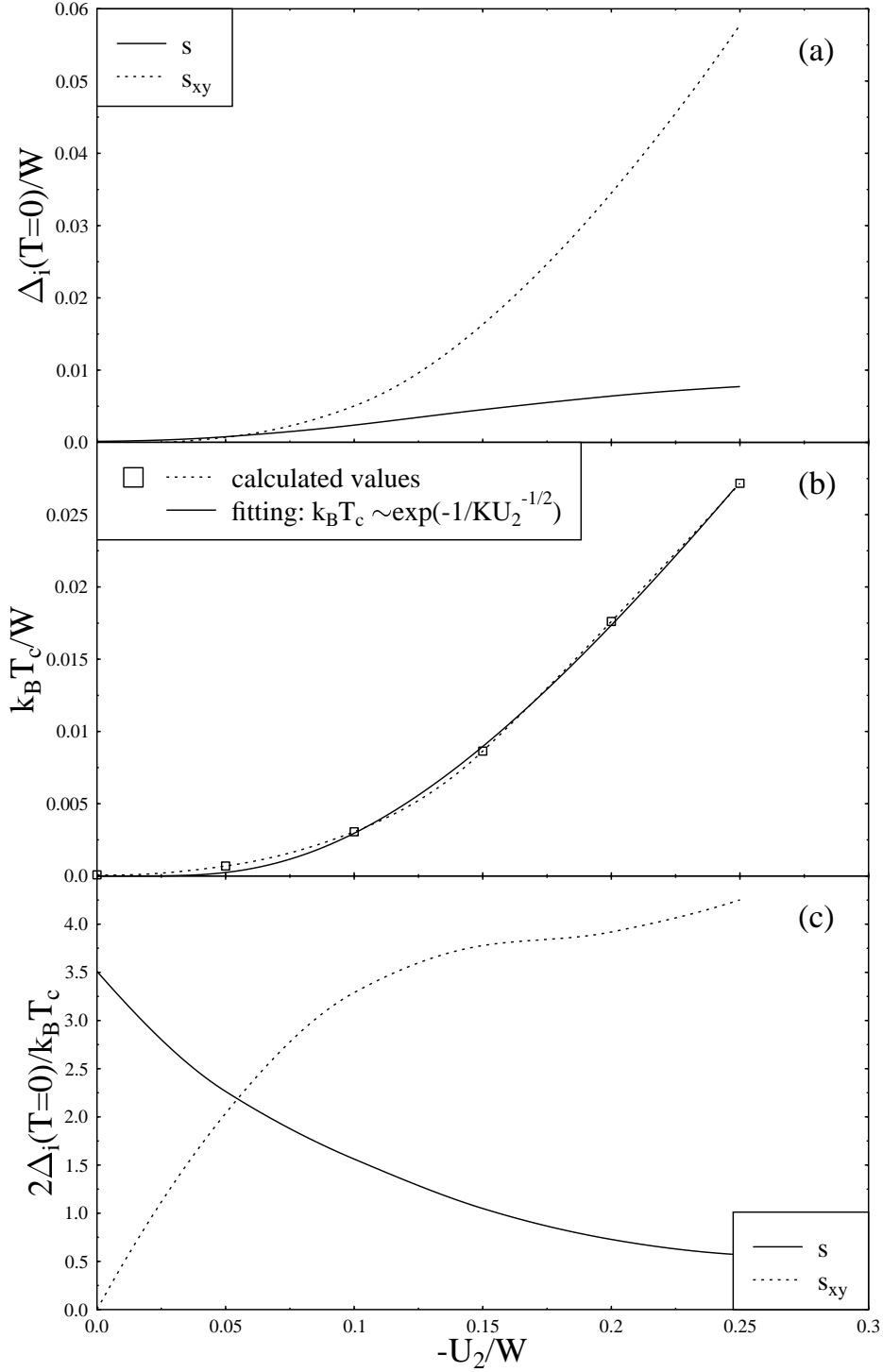, width=13cm, angle=180, bbllx=140, bblly=55,
  bburx=520, bbury=630}
\caption{The effect of $U_{2}$ on components of a stable $A_{1}$-symmetry
solution at $U_{0}\!=\!-0.1W$, $U_{1}\!=\!0.0$ at half-filled band; plot 
(a) shows the zero temperature gap amplitude versus $U_{2}$ dependence,
plot (b) depicts $U_{2}$ dependence of the critical temperature while 
plot (c) illustrates $U_{2}$ dependence of the ratio
$2\Delta(T\!=\!0)/k_{B}T_{c}$.}
\label{fig8}
\end{center}
\end{figure}

In Fig.\ \ref{fig8} we focus our attention on the effects accentuately caused 
by {\em NNN} interaction terms in $T_c$, in zero temperature gap-amplitude and
in the ratio of $R\!=\!2\Delta_{0}/k_{B}T_{c}$. For simplicity, $U_{1}$ is 
set to be zero in this study. This particular choice results in an order 
parameter of the form $\Delta(\vec{k})\!=\!\Delta_{s} \!+\!
\Delta_{s_{xy}}\cos{k_{x}}\!\cos{k_{y}}$, (i.e. only {\em s} and $s_{xy}$ 
components). The values of $U_{0}$ and $U_{2}$ are tuned so that the
emerging $A_{1}$-symmetry pairing state becomes the stable superconducting 
state for the chosen set of interaction parameters. The effect of $U_{2}$ on 
the mentioned quantities is presented in the figure. We note, that for 
$U_{1}\!\ne\!0.0$ values a similar analysis can be done leading essentially
to the same features.
 
First of all, Fig.\ \ref{fig8}(a) shows the $U_{2}$ dependence of the gap 
amplitudes at zero temperatures. One can immediately realize the
exponential type behaviour of $\Delta_{s_{xy}}$ versus $-U_{2}$ for
small, $|U_2|/W \!\ll\! 1.0$ values. At the same time, $\Delta_{s}$ has 
a linear dependence on $-U_{2}$. The $U_{2}$ dependence of the critical 
temperature is exemplified in Fig.\ \ref{fig8}(b) for the $A_{1}$-symmetry 
solution. A good fit to the numerically calculated 
points can be obtained by the $T_{c} \sim \exp{\![-1/(K\sqrt{U_{2}})]}$ 
relation, where $K \!=\! 0.5226$. Such type of $T_c$ behaviour related to 
non-on-site interactions is not unusual. For example, a similar 
relation was suggested by Dahm {\it et al.} (Dahm {\em et al.}) for 
the nearest neighbour coupling dependence of $T_{c}$, instead of an 
$\exp{\![-1/(\tilde{K}g_{i})]}$ functional form. Concerning the ratio of the 
zero temperature gap amplitude to $T_c$, in Fig.\ \ref{fig8}(c) the 
$R \!=\! 2\Delta_{0}/k_{B}T_{c}$ values are plotted versus $-U_{2}/W$ for the 
two components of the $A_{1}$ solution. In the $U_{2} \!=\! 0$ case (where the 
order parameter has the form of $\Delta(\vec{k})\!=\!\Delta_{s}$) this ratio 
is about $3.52$ and hence corresponds to the {\it BCS} predicted
value. However, if {\em NNN} interaction is turned on an $s_{xy}$
component of the order parameter 
develops rapidly resulting in an increase (decrease) of the ratio {\em R} for 
the $s_{xy}$ ({\em s}) channel.

Based on these results it can be seen that nonzero next-nearest neighbour  
coupling, besides the emergence of different possible superconducting species,
also influences the main characteristic properties of the
superconducting phase.  

\section{Summary and Conclusions}
\label{concl}

The superconducting phase diagram of the two-dimensional extended Hubbard 
model containing hopping and interaction terms with spatial dependences 
exceeding nearest neighbour distance in range has been systematically analyzed
in mean-field approximation. The possible superconducting phases of different
symmetry were taken into account unrestrictedly. For building up the phase 
diagrams, the emerging phase was choosen on the basis of a free-energy (or at
$T\!=\!0$ ground state energy) analysis of all the possible solutions of the gap 
equations for every fixed set of coupling constants, temperature and doping 
values. Main characteristics of different superconducting phases including 
critical temperatures $T_{c}$, zero temperature gap amplitudes $\Delta_{0}$, 
$\Delta_{0}/T_c$ ratios, temperature and doping dependences were also studied in detail for 
every domain of the phase diagram. The obtained results clearly underline the 
importance of next nearest neighbour terms (both hopping and interaction
contributions) in determining the prominent superconducting properties of the 
system.

The present paper explored such a parameter domain of the phase diagram which 
has not been investigated so far systematically in literature starting
with Fig.\ \ref{fig1}, where in the presence of only nearest neighbour terms, 
non half-filled band case was also investigated. Starting from this level 
different next-nearest neighbour contributions were taken gradually into 
consideration within the extended Hubbard Hamiltonian, their effects being
systematically presented in Figs.\ \ref{fig2}-\ref{fig8} and analyzed in 
detail with physical implications being emphasized, especially those connected 
to the presence of next-nearest neighbour terms. 

On this line the following interesting features deserve attention:
double phase transitions with decreasing temperature (following the emergence 
of a superconducting state at $T_c^{(1)}$ the symmetry of the superconducting
order parameter changes at $T_c^{(2)} \!<\! T_c^{(1)}$ in a first-order phase 
transition); quantum phase transitions at zero temperature between pairing 
states of different symmetry driven by doping; enhancement in the 
resistance of the stable superconducting phase against thermal pair-breaking 
effects in the presence of interaction terms exceeding nearest neighbour 
distance in range; increase in the $T\!=\!0$ gap-amplitude over $T_{c}$ ratio 
for some symmetry species; the main superconducting properties become more 
sensible with respect to doping; strong asymmetry between hole and electron 
doped cases; rich spectrum of stable superconducting states of different 
symmetry (in the presence of repulsive on-site and in some cases even for 
repulsive nearest neighbour interactions, the $A_{1}$ symmetry is favored for 
electron, and a $d$-wave type order parameter for hole doping values); 
elimination of {\bf k}-independent order parameters even at half-filling, 
and so on.

We strongly hope, that the presented results will constitute a valuable
starting point for further theoretical investigations related to the effect
of long-range contributions in building up superconducting properties of
the system under study. 
 
\section{Acknowledgements}

One of the authors (Zs.Sz.) would like to acknowledge earlier financial
support by SOROS Foundation and present financial supports of the 
Universitas and the Pro Regione Foundations of Kossuth Lajos University. 
For Zs.G. research was supported by the Hungarian National Science Funds
under grant OTKA-T013952. Furthermore, both authors thank financial support 
of the Research Group in Physics of the Hungarian Academy of Sciences at 
Kossuth Lajos University, Debrecen.

\section{References}
\noindent P. W. Anderson, Science {\bf 235} (1987) 1196. \\
E. Dagotto, Rev. Mod. Phys. {\bf 66} (1994) 763. \\
M. G. Smith, A. Manthiram, J. Zhou, J. B. Goodeneough and 
J. J. Market, Nature {\bf 351} (1991) 549. \\
R. Micnas, J. Ranninger, and S. Robaszkiewicz, Rev. Mod. Phys.
{\bf 62} (1990) 113. \\
See for example Proc. Inter. Conf. on Strongly Correlated
Electron Systems, Amsterdam 1994, SCES-94, edited by F. R. de Boer, P. F. de 
Chatel, J. J. M. Franse and A. de Visser, North-Holland, Elsevier, 1995.\\
J. de Boer, V. E. Korepin  and A. Schadschneider,
Phys. Rev. Lett. {\bf 74} (1995) 789.\\
T. Koma and H. Tasaki, Phys. Rev. Lett. {\bf 68} (1992) 3248. \\
H. Tasaki, Phys. Rev. Lett. {\bf 75} (1995) 4678. \\
J. A. Verges, F. Guinea, J. Galan, P. G. J. van Dongen,
G. Chiappe and E. Louis, Phys. Rev. {\bf B49} (1994) 15400. \\
A. F. Veilleux, A. M. Dar\'e, L. Chen, Y. M. Vilk and
A. M. S. Tremblay, Phys. Rev. {\bf B52} (1995) 16255. \\
M. Lavagna and G. Stemmann, Phys. Rev. {\bf B49} (1994) 4235.\\
P. B. Littlewood, J. Zaanen, G. Aeppli and H. Monien, Phys. Rev. 
{\bf B48} (1993) 487. \\
G. Blumberg, B. S. Stojkovic and M. V. Klein, Phys. Rev. 
{\bf B52} (1995) R15741. \\
M. Gul\'acsi, A. R. Bishop and Zs. Gul\'acsi, Physica {\bf C244} (1995)
87. \\ 
A. A. Abrikosov, Phys. Rev. {\bf B51} (1995) 11955 and ibid.,
{\bf B52} (1995) R15738, Physica {\bf C222} (1994) 191 and ibid.,
{\bf C233} (1994) 102. \\
V. Ponnambalam and U. V. Varadaraju, Phys. Rev. {\bf B52} (1995) 16213.\\
R. Fehrenbacher and M. R. Norman, Phys. Rev. Lett. {\bf 74} (1995) 3884.\\
C. O'Donovan and J. P. Carbotte, Phys. Rev. {\bf B52} (1995) 16208. \\
W. Brenig, Physics Reports {\bf 251} (1995) 155.\\
R. Strack and D. Vollhardt, Phys. Rev. Lett. {\bf 72}
  (1994)  3425.\\
R. Strack and D. Vollhardt, Jour. Low Temp. Phys. {\bf 99} (1995) 385.\\
C. Verdozzi and M. Cini, Phys. Rev. {\bf B51} (1995) 7412.\\
C. Verdozzi, in NATO Advanced Study Institute Series B, 
Physics, 1993, Kluwer and Dordrecht, page 237.\\
D. M. King et al. Phys. Rev. Lett. {\bf 70} (1993) 3159 \\
M. Di Stasio and X. Zotos, Phys. Rev. Lett. {\bf 74} 2050, (1995).\\
J. van den Brink, M. B. J. Meinders, J. Lorenzana, R. Eder and
G. A. Sawatzky, Phys. Rev. Lett. {\bf 75} (1995) 4658.\\
G. Grigelionis, E. E. Tornau and A. Rosengren, Phys. Rev. 
{\bf B53} (1996) 425. \\
A. P. Kampf, Physics Reports, {\bf 249} (1994) 222. \\
A. A. Abrikosov, L. P. Gor'kov and I. E. Dzyaloshinskii, 
in Methods of Quantum Field Theory in Statistical Physics ({\it Pergamon 
Press Ltd.}, 1965).\\ 
T. Dahm, J. Erdmenger, K. Schanberg and C. T. Rieck, 
Phys. Rev. {\bf B48} (1993) 3896. \\
B. L. Gyorffy, J. B. Staunton and G. M. Stocks, 
Phys. Rev. {\bf B44} (1991) 5190.\\
W.H Press, S.A. Teukolsky, W.T. Vetterling and 
B.P. Flannery: Numerical Recipes in Fortran ({\it Cambridge University
Press}, 2nd edition, 1992). \\
R. Micnas, J. Ranninger, S. Robaszkiewicz and S. Tabor, 
Phys. Rev. {\bf B37} (1988) 9410.\\
R. Micnas, J. Ranninger and S. Robaszkiewicz, 
Phys. Rev. {\bf B39} (1989) 11653.\\
F. Wenger and S. \"Ostlund, Phys. Rev. {\bf B47} (1993) 5977.\\
S. Zhang, Phys. Rev. {\bf B42} (1990) 1012.\\
M. Mierzejewski and J. Zielinski, Phys. Rev. {\bf B52} (1995) 3079.\\
Y. M. Gufan, G. M. Vereshkov, P. Toledano, B. Mettout,
R. Bouzerar and V. Lorman, Phys. Rev. {\bf B51} (1995) 9228.\\
M. Inaba, H. Matsukawa, M. Saitoh and H. Fukuyama, Physica 
{\bf C257} (1996) 299.\\
D. J. Scalapino, Physics Reports {\bf 250} (1995) 331.\\
J. R. Iglesias, B. H. Bernhard and M. A. Gusmao, 
Physica {\bf B206-207} (1995) 678.\\
E. Cappelluti and L. Pietronero, Phys. Rev. {\bf B53} (1996) 932.\\
E. Dagotto, A. Nazarenko and A. Moreo, Phys. Rev. Lett. 
{\bf 74} (1995) 310.\\
P. Bourges, L. P. Regnault, Y. Sidis and C. Vettier, Phys. Rev.
{\bf B53}, (1996) 876.\\
K. Mashimoto, K. Nakao, H. Kado and N. Koshizuka, Phys. Rev. 
{\bf B53} (1996) 892.\\
V. E. Gasumyants, N. V. Ageev, E. V. Vladimirskaya, V. I. 
Smirnov, A. V. Kazanskiy and V. I. Kaydanov, Phys. Rev. {\bf B53}
(1996) 905.\\
R. Hopfengartner, M. Leghissa, G. Kreiselmeyer, B. Holzapfel,
P. Schmitt and G. S. Ischenko, Phys. Rev. {\bf B47} (1993) 5992.\\
B. Normand, H. Kohno and H. Fukuyama, Phys. Rev. {\bf B53} (1996) 856.\\

\end{document}